\documentclass[aps,prl,showpacs,twocolumn,nofootnoteinbib,groupedaddress]{revtex4}
\usepackage{epsfig,amssymb,amsmath,latexsym}
\usepackage{verbatim}
\usepackage{xcolor}   
\usepackage{amsfonts}
\usepackage{amsmath}
\usepackage{amssymb}
\usepackage{amsthm}
\usepackage{hyperref} 
\usepackage{wasysym}
\usepackage{bm} 
\usepackage{physics}

\begin{document}
\textheight=23.8cm
\title{\Large  Dissipation and quantum noise in chiral circuitry}
\author{Disha Wadhawan}
\affiliation{Department of Physics and Astrophysics, University of Delhi, Delhi 110007}
\author{Sourin Das}
\affiliation{Department of Physical Sciences, IISER Kolkata, Mohanpur, West Bengal 741246}
\affiliation{Department of Physics and Astrophysics, University of Delhi, Delhi 110007}

\email{disha1_wadhawan@yahoo.co.in, sdas@physics.du.ac.in}
\date{\today}
\date{\today}
%
%
\begin{abstract}
We obtain an empirical out of equilibrium fluctuation-dissipation type relation pertaining to a linear relation between dissipation and quantum shot noise generated by a quantum point contact (QPC) which  induces inter-edge scattering of electrons among ``$n$'' number of chiral edges of $\nu=1$ quantum Hall state where ``$n$'' is an integer ($n\geq 2$). We have shown that the ratio of total maximum power dissipated at the QPC ($D_{total}$) to the sum of auto-correlated noise  generated  in the chiral edge channels emanating out of the QPC region ($S_{total}$)  is given by  $D_{total}/S_{total}{\small{(}}\omega=0{\small{)}} = V/{\small{}}4  e{\small{}}$ where  $e$ is the electronic charge and $V$ is the voltage imposed on any one of the ``$n$'' incoming edge channels while keeping remaining ``$n-1$'' edge channels grounded. This implies that this ratio is universal except for a linear voltage bias dependence, i.e., it is independent of details of the scattering matrix ($S$-matrix) of the QPC region. Here the maximum power dissipation in each chiral edge is defined as the rate at which energy would be lost if the non-equilibrium distribution of electrons generated by the QPC region in each chiral edge is equilibrated to the corresponding zero temperature Fermi distribution. Further,  for ${\cal Z}_n$ symmetric $S$-matrix,  we show that the universal behaviour persists even when all the bias voltages imposed on the incoming edge channels are kept finite and distinct.

\end{abstract}
%
\maketitle
\begin{figure}
\centering
\includegraphics[width=.6\columnwidth]{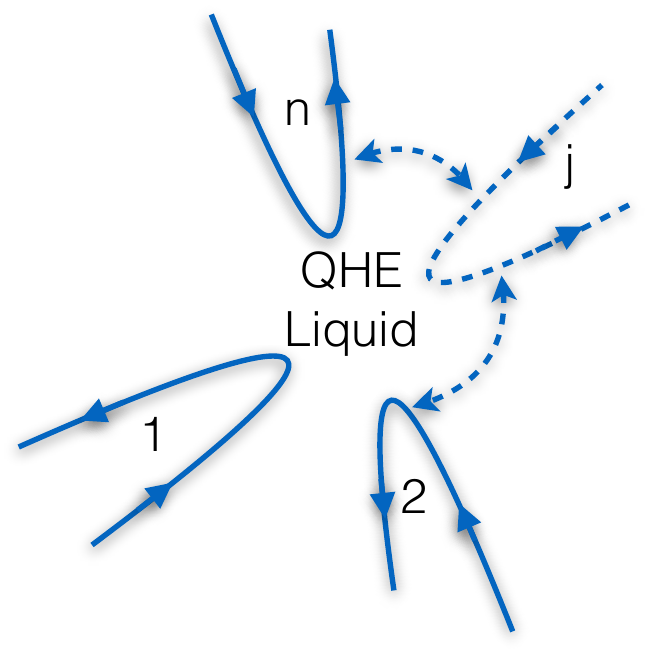}
\caption{The central part of the figure represents a complex QPC which has ``$n$'' number of QH edge states meeting at the QPC region. The opposite arrows represent the chiralities of the edge states.}
\label{nterminal}
\end{figure}
\begin{figure}
\centering
\includegraphics[width=.85\columnwidth]{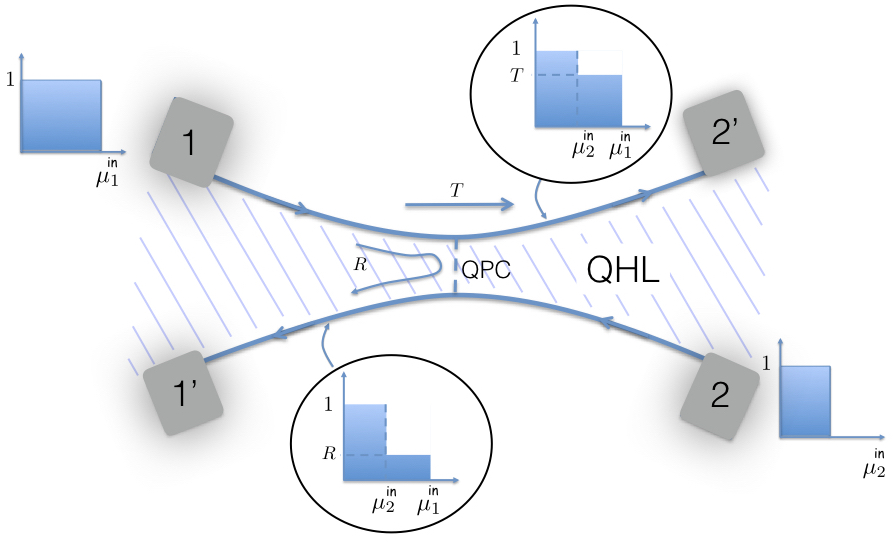}
\caption{A two terminal set-up is shown where, $\mu_{1}^{in}$ and $\mu_{2}^{in}$ are the chemical potentials of the electron reservoirs feeding the two incoming edge states respectively. The  regions of contact between the edge states and the electron reservoirs $1,1'$ and $2, 2'$ are considered to be far from the QPC region. The  chemical potentials  $\mu_{1}^{in}$ and $\mu_{2}^{in}$ of the incoming edge states are related to the voltage applied on the leads $1, 2$ as $V_1=e \mu_{1}^{in}$ and $V_2=e \mu_{2}^{in}$ where $e$ represents the electronic charge. }
\label{2terminal}
\end{figure}
\begin{figure}
\centering
\includegraphics[width=0.75\columnwidth]{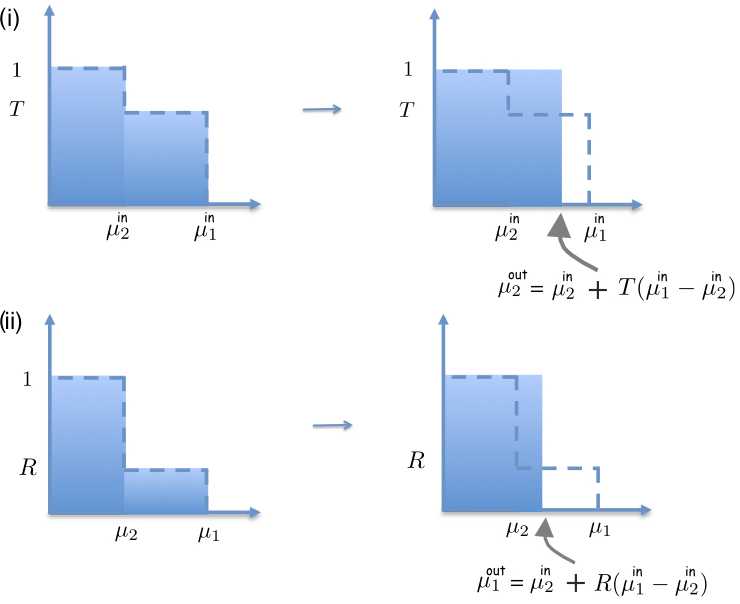}
\caption{Non-equilibrium energy distribution of  the electrons (on left hand side of the figure)  in the outgoing channel and their corresponding equilibrated distribution (on right hand side of the figure) are depicted for (i) the transmitted edge and (ii) reflected edge. Note that here $R=|r|^{2}$ and $T=|t|^{2}$.}
\label{equilibration}
\end{figure}

{{\emph{Introduction :}}} Measurement of $dc$ shot noise\cite{Beenakker, Blanter} of the electronic current has served as a leading probe for studying quantum Hall (QH) edge states\cite{Halperin}. Such noise measurements have been extensively used for investigating exotic physics\cite{stern} associated with the QH effect. A classic example of this fact is the detection of quasi-particle excitations with fractional charge\cite{Moty,Etienne,Glattli,Hashisaka} via shot noise measurements. Also, shot noise measurements of electronic current in solid state setting are known to be extremely valuable in probing fundamental aspects of quantum mechanics like the principle of complementarity\cite{YuvalMoty}, controlled study of the phenomenon of dephasing\cite{MotyDephasing} as well as probing of quantum statistics via phenomenon of bunching or anti-bunching\cite{Tarucha}. Studies like detection of quantized flow of heat carried by electrons\cite{Jezouin,Sothmann} and  anyons\cite{MotThermalQ} in the QH edge states have been successful with the help of shot noise measurements. Interestingly, the QH edge states also provide a lucrative platform for studying physics of electrons driven out of equilibrium and related equilibration and dissipation in a simplest  possible setting\cite{edge,IQHR,QHchannels,PhysRevB.84.085105,PhysRevLett.109.106403} by virtue of their chiral nature and topological protection by a bulk energy gap.\\
In this backdrop, we explore the possibility of obtaining a universal ratio of power dissipation to the corresponding zero temperature, zero frequency quantum noise for a multi-terminal chiral circuitry. In particular, we consider a QPC\cite{QPC})  geometry which can scatter electrons between $n$ chiral edge channels of $\nu=1$ QH state. The $n$ edge channels which are flowing into the scattering region  are called incoming edge  channels and those which are emanating out of the region of scattering are called outgoing edge channels (see Fig.~\ref{nterminal}). \\
We define the maximum power dissipation as the difference between the sum of power (energy flowing per unit time) associated with the non-equilibrium distribution of electrons on the outgoing edge channels and the corresponding equilibrated zero temperature Fermi distribution such that the particle density on each edge before and after equilibration remains same. This is nothing but the maximum amount of energy that can be extracted from the non-equilibrium distribution of electrons in the  outgoing edge per unit time. We show that sum of auto-correlated $dc$ shot noise\cite{Blanter} generated in each outgoing edge channel is linearly proportional to the total dissipation of the chiral circuitry as defined above. Additionally, we show that the proportionality constant is universal except for a dependence on the applied voltage bias. This provides an empirical, out of equilibrium fluctuation-dissipation  type of relation for free chiral fermions. Note that $S_{total}$ is a routinely measured quantity in QH experiments and we argue that $D_{total}$ is also an independently measurable quantity and hence a relation between these two quantities can have interesting  consequences. For instance, an observation of deviation from our empirical relation in QH edge in a QPC geometry in the very low temperature limit could be considered as a signature of deviation from Fermi liquid behaviour for integer QH edge states. \\
To illustrate this, let us consider the current-voltage characteristics reported in Ref.~\onlinecite{Roddaro} for $\nu=1$ edge state which strongly predicts a non-Fermi liquid (Luttinger liquid) behaviour. Note that, this conclusion of Luttinger liquid behaviour is based on the measurement of differential conductance. Another independent approach could be to measure quantum shot noise and the related Fano factor and look for signature of fractional charge as done in Ref.~\onlinecite{Hashisaka}. Alternatively an experimental study of deviation from our empirical relation could suggest yet another independent route for confirmation of a deviation from Fermi liquid behaviour which is different from both of the above mentioned approaches but merges elements of both (a combination of both noise and conductance measurement). 
The article is organised as follows: \\
\begin{enumerate}
\item[({\it {a}})] we start by evaluating power dissipation for the $n=2$ case which corresponds to the standard QPC geometry\cite{QPC} (see Fig.~\ref{2terminal}). By comparing total power dissipation with the sum total of zero temperature auto-correlated $dc$ shot noise in the two outgoing channels we obtain a linear relation of the form $D_{total}/S_{total}{\small{(}}\omega=0{\small{)}} = V/{\small{}}4  e{\small{}}$. Here $V$ is the voltage bias applied across the QPC. \\
\item[
({\it{b}})] next, we check if this is a generic feature of such a network of edge state and we evaluate this ratio for the general case of $n$ edges.  We find that the ratio remains universal and is given by $D_{total}/S_{total}=V/4e$ just like in the case of $n=2$ provided a voltage $V$ is imposed on one of the $``n"$ incoming edge channels while all other $``n-1"$ edge channels are kept grounded. \\
\item
[({\it{c}})] further, we consider a situation where finite and distinct voltages are imposed on the incoming channels. We show that in this case the ratio $D_{total}/S_{total}{\small{(}}\omega=0{\small{)}}$ is no more universal unless some symmetries are imposed on the $S$-matrix describing the junction. In particular, for the case of $n=3$ we show that if the $S$-matrix corresponding to the QPC has a ${\cal Z}_3$ symmetry, i.e., it is symmetric under cyclic permutation of the incoming edge channels, then the ratio is again given by $D_{total}/S_{total}=(1/4e) F(V_1,V_2,V_3)$ where  $ F(V_1,V_2,V_3)$ is only a function of  voltages $V_1,V_2$ and $V_3$ imposed on the three incoming edge channels and is independent of the $S$-matrix of the junction.\\
\item[({\it{d}})] finally, we show that it is not enough to have a ${\cal Z}_n$ symmetric $S$-matrix for obtaining a universal ratio for $n>3$. At this point we identify a geometric way of visualizing the possibilities for having universal ratio in terms of simple geometric figures. \\
\end{enumerate}%
\noindent{\underline{{\emph{Dissipation and noise for $n=2$ case :}}}} We start by discussing the case of dissipation and noise for $n=2$ which corresponds to the standard single QPC geometry depicted in Fig.~{\ref{2terminal}}, where we have two incoming edge channels propagating towards the QPC region and two outgoing edge channels moving out of the QPC region. We assume that the reservoir of electrons which are feeding the incoming edge channels with electrons make an ideal contact \cite{idealcontacts} with the edge state. We also assume that the reservoirs are maintained at zero temperature. The  reservoir $1 \& 2$ feeding the incoming edge channels (`incoming' refers to edge propagating towards the QPC) are maintained at chemical potential $\mu_{1}^{in} \& \mu_{2}^{in}$ respectively. 
Hence, the corresponding incoming current in each of these channels are given by the Hall relation: $(e^2/h)V_1$ and $(e^2/h)V_2$ where $e V_{1,2}= {\mu_{1,2}^{in}}$ ; $e$ is the electronic charge and $h$ is the Planck's constant. The distribution function of the incoming edge channels are given by the Fermi distribution at zero temperature, $f^{1,2}=\Theta(E-\mu_{1,2}^{in})$ respectively, where $E$ is the energy of the electron measured for an arbitrary chosen zero of energy. \\
Now, assuming that the $S$-matrix at the junction is given by $\{(s_{11'},s_{12'}),(s_{21'},s_{2 2'})\}$ such that $|s_{11'}|=|s_{22'}| = r$ and $|s_{12'}|=|s_{21'}| =t$, the distribution function for the outgoing edge channels (here `outgoing' refers to edge propagating away from the QPC) heading towards reservoir $1'$ and $2'$ are given by $f_{non-equ.}^{1}=|r|^2 \Theta(E-\mu_1^{in})+ |t|^2 \Theta(E-\mu_2^{in})$ and  $f_{non-equ.}^{2}=|r|^2 \Theta(E-\mu_2^{in})+ |t|^2 \Theta(E-\mu_1^{in})$ respectively. 
If we equilibrate the distribution (as shown in Figure \ref{equilibration}) corresponding to $f_{non-equ.}^{1,2}$ to a zero temperature Fermi distribution such that the electron density on the edge channel remains the same, the resulting new distribution is given by $f_{equ.}^{1,2}=\Theta(E-\mu_{1,2}^{out})$ where $\mu_{1}^{out}=|r|^2 \mu_1^{in}+ |t|^2 \mu_2^{in}$ and $\mu_{2}^{out}=|r|^2 \mu_2^{in}+ |t|^2 \mu_1^{in}$ respectively. The effective chemical potentials for the outgoing edge states given by $\mu_{1,2}^{out}$ can be extracted straightforwardly either by performing an integration of the non equilibrium distribution function and then equating the obtained result to integration of a zero temperature Fermi distribution or by identifying the currents that are injected into the outgoing edge channels and use the Hall relation to convert the currents into equivalent voltages which leads to the identification of the respective effective chemical potentials. 

Hence, the maximum power that can be extracted from these non-equilibrium distribution $f_{non-equ.}^{1,2}$ is given by the energy that will be released per unit time if we equilibrate these distributions to corresponding zero temperature Fermi distribution given by $f_{equ.}^{1,2}$. We call this quantity power dissipation and denote it by $D$. Hence, the power dissipation for each $i^{th}$ outgoing edge channel can be defined as 
\begin{equation}
D_{i}=\frac{1}{h}\int_{-\infty}^{\infty} \left\{ \, f_{non-equ.}^{i} -  f_{equ.}^{i} \, \right\} E dE .
\label{equ1}
\end{equation}
Here, we work with a linear dispersion model for the chiral edge channel given by $E_k= v_F\, \hbar \, k$, 
where $k$ is the momentum and $v_{F}$ is the fermi velocity of the electron on the edge. Also we assume that the bias window is small  enough so that the $S$-matrix can be considered to be independent of energy in this window. Hence $D_{i=1,2}$ can be straightforwardly evaluated to give  
\begin{equation}     
\begin{aligned}
D_{i=1}=D_{i=2}&=& &\frac{e^{2}}{2 h} |s_{1 2'}|^2\,|s_{1 1'}|^2  ~(V_1-V_2)^2& \\
                          &=& &\frac{e^{2}}{2 h} |s_{2 1'}|^2\,|s_{2 2'}|^2  ~(V_2-V_1)^2& \\
				&=& &\frac{e^{2}}{2 h} |r|^2\,|t|^2  ~(V_2-V_1)^2& 
\end{aligned}
\label{equ2}
\end{equation}

Now we will analyse the above result from the perspective of standard definition of power dissipation in a two terminal geometry given by  $P= I (V_2-V_1)$ where the net current $I$ flowing across the QPC is given by the Landauer formula, i.e., $I=(e^2/h) |t|^2 (V_2-V_1)$. For the two terminal set-up, the total resistance in the circuit can be visualised as a series combination of two resistances, {\it{(a)}} the contact resistance between the edge states and the electron reservoirs ($R_C$) and {\it{(b)}} the resistance due to scattering at the QPC region ($R_S$). Hence the total power dissipation in given by $P=I (V_2-V_1)= I^2 R_{total} = I^2 (R_C + R_S)$. Now, the point to note here is that the $D_{total}=D_1 + D_2$ corresponds to power dissipation owing to $R_S$ alone, i.e., $D_{total}= I^2 R_S$. This can be understood by rewriting $D_{total}$ in the form of $I^2 R_S$ which allows us to extract an expression of $R_S$ given by $R_S=(e^2/h)^{-1} (\vert r \vert^2/\vert t \vert^2 )$. This expression of $R_S$ can be readily identified with the four-terminal resistance of a quantum scatterer in a single one dimensional channel connected to ideal leads\cite{Imry1986, Y.Imry} which is nothing but the resistance of the quantum scatterer alone. Additionally, we would like to point out that dissipation due to a resistor is intimately related to Joule heating\cite{Benenti2017, Butcher1990, Anderson, Imry1986} and is a topic of great interest in nanoscopic and mesoscopic systems though this is not the focus of our work. 

It is interesting to note that though the non-equilibrium distributions ($f_{non-equ.}^{1,2}$) for the two outgoing edge channels are different but the dissipations $D_1$ and $D_2$ are identical. We will see later that this is valid only for the simplest case of a QPC ($n=2$). For the general case of $n$-terminal set-up the dissipation on each outgoing chiral channel are different from one another.  Also note that the dissipation is maximum when the transmission probability $|t|^2$ is exactly half for n=2. 

Since we are interested only in the total dissipation $D_{total}$ associated with the entire set-up and not the dissipation associated with each of the individual outgoing edge channel ($D_i$) therefore we do not  need information about the non-equilibrium distribution on each of the  $i^{th}$ outgoing edge channel and hence the total dissipation associated with the QPC can be expressed as 
\begin{equation}
D_{total}=\sum_{i} D_{i}= \frac{1}{{2h}} {\Big(}\sum ( \mu_i^{in})^2-(\mu_{i}^{out})^2{\Big)}
\label{equ3}
\end{equation}

It should be noted that the scattering at the QPC is elastic thus, the total energy carried by the incoming edge channels with equilibrium distribution function for electrons and the outgoing edge channel with the non-equilibrium distribution function are the same, i.e., 
\begin{equation}
\frac{1}{2 h} \sum_{i}  (\mu_i^{in})^2=\frac{1}{h} \sum_{i} \int_{-\infty}^{\infty} f_{non-equ.}^{i} \hspace{0.08cm}E\hspace{0.08cm}dE
\label{equ4}
\end{equation}
Hence, substituting Eq.~(\ref{equ4}) in Eq.~(\ref{equ1}) and performing the integral directly provides Eq.~(\ref{equ3}) as expected. But while working with Eq.~(\ref{equ3}) one must keep in mind that the power dissipation for each individual outgoing edge channel  $D_i$ as defined in Eq.~(\ref{equ1}) is not equal to $(1/2h) [(\mu_i^{in})^2-(\mu_{i}^{out})^2]$. Finally, we take a note of the fact that the total dissipation is actually easily accessible experimentally in the following sense: the right hand side of Eq.~(\ref{equ3}) comprises of $\mu_{in}$'s and $\mu_{out}$'s. The $\mu_{in}$'s correspond to voltages that we impose on the incoming edge channel and hence are known quantities while the $\mu_{out}$'s are nothing but the read outs of an ideal voltage probe \cite{Y.Imry} implemented on the outgoing edge channels. This makes the total power dissipation an experimentally measurable quantity though the power dissipation in each outgoing channel can not be accessed using the same voltage probe measurement. In principle, recent experimental advances allow for direct measurement of the non-equilibrium distribution function of electrons on the edge channels  by using tunnel coupled quantum dots \cite{QHchannels} which could also facilitate the indirect measurement of $D_i$ on each individual outgoing edge channel. But this calls for a rather complicated experimental protocol. 

An efficient way to express the total power dissipation is to make use of the current splitting matrix\cite{amit}. For the QPC geometry the current splitting matrix connect the incoming current vector given by  $\vec{I}_{in}=(e/h) \, [\, \mu_1^{in},\, \mu_2^{in} \,]$ to the outgoing current vector  $\vec{I}_{out}=(e/h) \, [\, \mu_1^{out}, \, \mu_2^{out} \,]$  via relation 
\begin{equation}
\vec{I}_{out}= M  \cdot \vec{I}_{in},
\label{equ5}
\end{equation}
where, $M$ is the current splitting matrix which is formed by replacing each element of the S-matrix by its corresponding squared modulus. Hence, for the case of a single QPC, it is given by M = $[\, \{ \, |r|^2,\,|t|^2\},\,\{ \, |t|^2,\,|{r}|^2 \, \} \,]$.  Later we will see that the ratio $D_{total}/S_{total}$ can be expressed solely in terms of elements of the $M$ matrix which is the primary motivation for introducing this matrix. 
Now, we turn to the discussion of zero frequency quantum noise which is popularly known as \emph{dc} shot noise at zero temperature. This quantity provides information about the temporal correlations in electron transport which cannot be obtained from conductance measurements  alone \cite{Buttiker1990,1996cond.mat.11140D,Blanter}.
In this context, the noise power is defined as the Fourier transform of the correlation between the fluctuations in average current in contact $i$ and the fluctuation in average  current in contact $j$, which at zero frequency and zero temperature\cite{Blanter} is given by 
\begin{equation}
\begin{split}
S_{\alpha\beta}=\dfrac{e^{2}}{h}\sum_{\gamma\neq\delta}\int dE (s_{\alpha\gamma}^{\dagger}s_{\alpha\delta}s_{\beta\delta}^{\dagger}s_{\beta\gamma})\left\{f_{\gamma}(E)[1-f_{\delta}(E)]\right.\\\left.+f_{\delta}(E)[1-f_{\gamma}(E)]\right \}  
\label{equ7}
\end{split}
\end{equation}
where $s_{\alpha\delta}$ represents elements of a $n \times n$ $S$-matrix describing the localized scatterer which facilitates scattering between $n$ edge channels and $f_{\alpha}$ represents Fermi distribution function associated with the $\alpha^{th}$ incoming edge channel. 
Hence the total auto-correlated noise generated by the point contact calculated at zero temperature using Eq.~\ref{equ7} is obtained as 
\begin{equation}
S_{auto}=S_{11}+S_{22}=\frac{2e^{3}}{h}\vert r\vert^{2}\vert t\vert^{2}\vert V_{1}-V_{2}\vert,
\label{equ9}
\end{equation}
and we find that $S_{auto}$ is related to $D_{total}$ (see Eq.\,\ref{equ2}) as  
\begin{equation}
S_{auto}=\left (\frac{4e}{V}\right )D_{total}, 
\label{equ10}
\end{equation}
where $V$ is the applied voltage bias across the junction. As argued earlier, $D_{total}$ is a measurable quantity and $S_{auto}$ is routinely measured in experiments \cite{Moty}. Thus, {Eq.~\ref{equ10} provides a direct relation between the two measurable quantities that can be experimentally tested.
Also, this relation between zero temperature noise and dissipated power at the point contact is one of the main results of our work that motivates us to examine existence of such a relation for $n>2$. \\
\noindent\underline{{{\emph{Dissipation and noise for a general $n$ :}}}} Now, we consider the general case of $n$-terminals ($n>2$) and investigate if such a relation holds. Using the expression of dissipation and noise given in Eq.~\ref{equ3} and Eq.~\ref{equ7} and expressing them solely in terms of elements of $M$-matrix we arrive at the following expression:
\begin{eqnarray}
S_{total}=2eG\sum_{\alpha}S_{\alpha\alpha}=e G\sum_{\alpha,\gamma,\delta \atop \gamma\neq\delta}M_{\alpha\gamma}M_{\alpha\delta}\vert V_{\gamma}-V_{\delta}\vert, \\
D_{total}=\sum_{i}D_{i}=\dfrac{G}{4}\sum_{\alpha,\gamma,\delta \atop \gamma\neq\delta}M_{\alpha\gamma}M_{\alpha\delta} [V_{\gamma}-V_{\delta}]^{2},
\end{eqnarray}
where $G=e^2/h$ and $S_{total}$ is the sum total of the auto-correlated noise. As we are interested only in the auto-correlated noise hence we are able to express $S_{total}$ solely in terms of the elements of current splitting matrix $M$  alone as the phases appearing in the S-matrix do not play any role here. Of course this will not be the case for the cross-correlated noise but that is not even of our interest for this article. An inspection of the above expression immediately reveals that the coefficient of each of the voltage difference $|V_{\gamma}-V_{\delta}|$ term appearing inside the summation in the expression for both the total noise and the total dissipation are identical. If we set all the voltages but one to zero then the ratio $D_{total}/S_{total}$ indeed becomes universal (independent of S-matrix) and is given by $V/4e$ where $V$ is that very voltage which is kept finite. Note that for a given $n$ the universality of the ratio $D_{total}/S_{total}$ can be checked against $n$ distinct cases where voltage is applied on one edge state at a time while all others are kept grounded. This fact could be of use to check the robustness of our result while exploring an experimental verification. This relation between $D_{total}$ and $S_{total}$ can be interpreted  as an empirical out of equilibrium fluctuation-dissipation relation for the free chiral fermion circuitry at zero temperature. This is the central result of this article. 
It is obvious from the expression for $S_{total}$ and $D_{total}$ given above that the ratio $D_{total}/S_{total}$ is in general not independent of the S-matrix. Next, it is a natural question to ask if other than the special case discussed above where all voltages are set to zero except one, can one expect a universal ratio. Now,  we will show that for certain symmetry consideration of the S-matrix for the junction can lead to universal ratios for $D_{total}/S_{total}$ even when all the voltage differences are kept finite and distinct. To see this we first rewrite the expressions for $D_{total}$ and $S_{total}$ as 
\begin{eqnarray}
S_{total}=e G~ \sum_{\gamma,\delta,\,{\gamma>\delta}} {\bra{\gamma}\ket{\delta}}  ~ \vert V_{\gamma}-V_{\delta}\vert~, \\
D_{total}=\frac{G}{4}~\sum_{\gamma, \delta,\,{\gamma>\delta}} {\bra{\gamma}\ket{\delta}} ~ [V_{\gamma}-V_{\delta}]^{2}~,
\end{eqnarray}
where $\ket{\gamma}, \ket{\delta}$ corresponds to the $N$ dimensional vectors in euclidean space with all real entires which are given by the $\gamma^{th}$ and $\delta^{th}$ column of the $M$-matrix. In order to demonstrate the usefulness of expressing $S_{total}$ and $D_{total}$ in terms of the inner product ${\bra{\gamma}\ket{\delta}}$ for obtaining conditions which will lead to universal value of $D_{total}/S_{total}$ we first consider the case of $n=3$ as discussed below.  

\noindent{\underline {{\emph{Dissipation and noise for the $n=3$ case:}}} } For $n=3$ we will now show that the ratio of $D_{total}$ to $S_{total}$ remains universal as long as we impose a S-matrix at the junction which respects ${\cal Z}_3$ symmetry. For a ${\cal Z}_3$ symmetric case, if we apply cyclic permutation of the labels of the incoming and the outgoing channels then it must keep the $S$-matrix invariant. Explicit construction of such a $S$-matrix for the case of $n=3$ can be found in Ref.~\cite{Chamon} which can be straightforwardly extended to the case of $n>3$. If the S-matrix has this symmetry then the corresponding $M$-matrix is ensured to have the same symmetry as its elements comprises of mod square of corresponding elements of the $S$-matrix. It is easy to check that a ${\cal Z}_3$ symmetric $S$-matrix will imply that ${\bra{1}\ket{2}}={\bra{2}\ket{3}} ={\bra{3}\ket{1}}$ where ${\ket{1}},{\ket{2}}$ and ${\ket{3}}$ are the first, second and the third column of the $M$-matrix respectively. This in turn implies that the coefficient of each of the three distinct voltage differences given by $\vert V_{1}-V_{2}\vert, \vert V_{2}-V_{3}\vert$ and $\vert V_{3}-V_{1}\vert$ appearing in the expression for $S_{total}$ and $D_{total}$ have identical values which leads to the ratio of $D_{total}/S_{total}$ to be given by 
\begin{eqnarray}
D_{total}/S_{total}= \frac{1}{4e} \frac{\sum^3_{\gamma,\delta=1,\,{\gamma>\delta}}\vert V_{\gamma}-V_{\delta}\vert^2}{ \sum^3_{\gamma,\delta=1,\,{\gamma>\delta}} \vert V_{\gamma}-V_{\delta}\vert}~,
\end{eqnarray}
which is clearly independent of the S-matrix describing the scattering of electrons between the three different edge states at the QPC and hence is universal within the constraint of ${\cal Z}_3$ symmetry. 
\begin{figure}
\centering
\includegraphics[width=1\columnwidth]{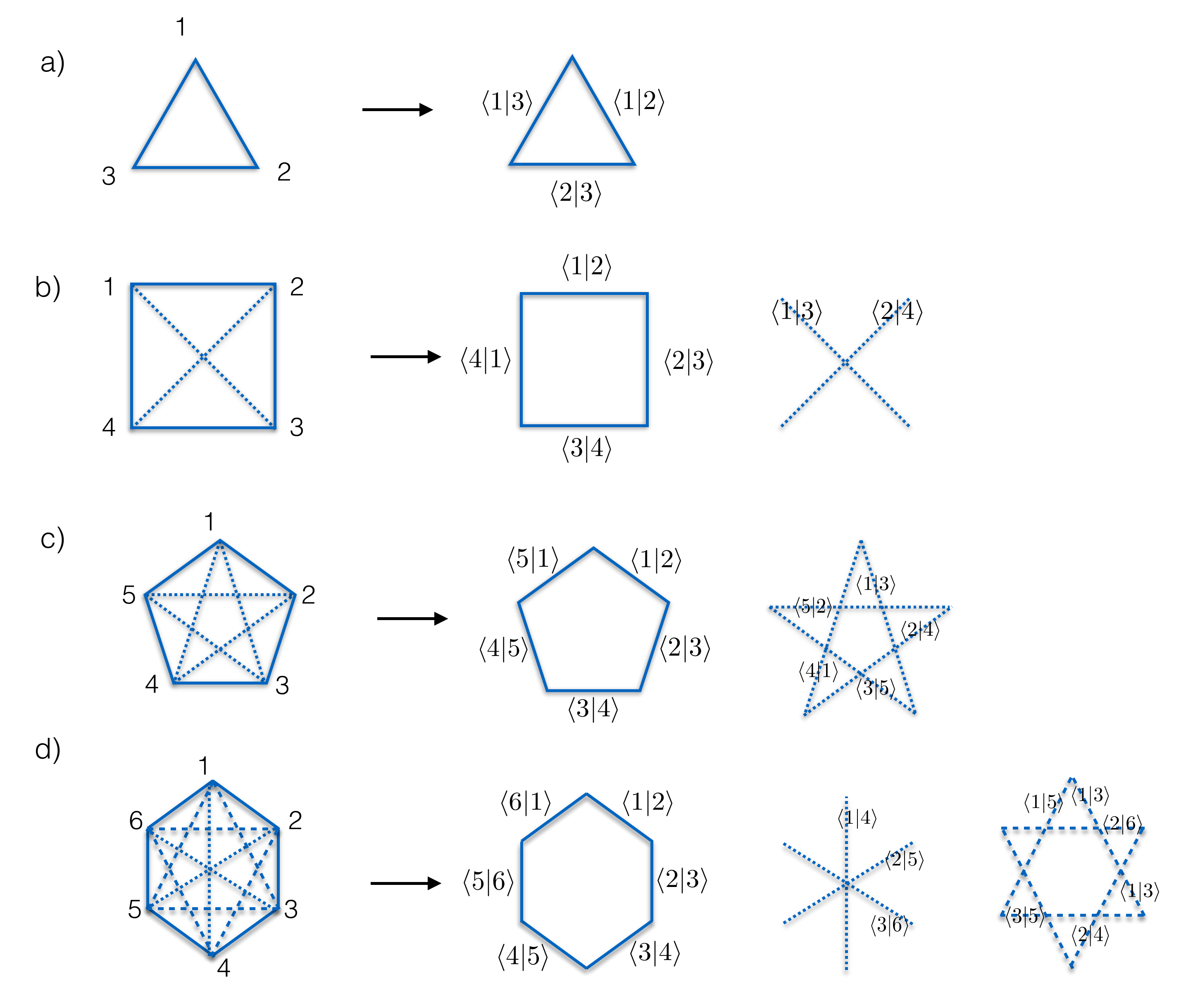}
\caption{On right hand side of figures: $a),b),c)$  and $d)$, geometric representations of all possible subsets of overlap which form a closed set under cyclic permutation ( under ${\cal Z}_n$) are shown for the case of $n=3,4,5,6$ respectively.}
\label{symmetry}
\end{figure}
\\
\noindent{\underline {{\emph{Dissipation and noise for $n\geq4$ case:}}} } Next, we would like to understand if ${\cal Z}_n$ symmetry for the $S$-matrix ensures such a universality for any given $n\geq4$. So, we consider the $n=4$ case next and try to understand if  ${\cal Z}_4$ symmetry for the $S$-matrix ensures such a universality. It is clear from the example of $n=3$, if due to some symmetry all the $\bra{\gamma}\ket{\delta}$ become equal, this will lead to a universal ratio of $D_{total}/S_{total}$ except for a voltage dependence. Hence the question of universal ratio for the $n=4$ case boils down to asking whether all possible distinct overlaps $\{ \bra{\gamma}\ket{\delta} \}$ for the $n=4$ case become equal if ${\cal Z}_4$  symmetry is imposed on the $S$-matrix. It is straight forward to check that this is not the case. But the interesting point lies in the fact that the full set of overlaps $\{ \bra{\gamma}\ket{\delta} \}$ splits into two distinct subsets where all elements of a given subset have same value. 

To obtain a geometric view of existence of these subset we start with the following construction. We assign the index $\gamma=1,2,3,4$ to the four edge state which are meeting each other at the QPC region for $n=4$ case (see Fig.~\ref{nterminal}). We map these four distinct indices to four distinct points (call it vertices, see Fig.~\ref{symmetry} b)) organized as the vertex of a regular polygon and then we join these point with all possible line (call it bonds). We then note that there are two kinds of bonds which are distinct. First ones are of the kind which form the sides of a polygon of 4 sides and second one form the diagonals of the polygon.We can assign each of these bonds a value given by $\bra{\gamma}\ket{\delta}$ where this bond corresponds to the line joining $\gamma^{th}$ and the $\delta^{th}$ vertex as shown in Fig.~\ref{symmetry} b). The interesting consequence of imposition of $z_4$ symmetry lies in the fact that all the overlaps belonging to a given kind of bond (sides or diagonal) have to be equal, i.e.,  ${\bra{1}\ket{2}}={\bra{2}\ket{3}}={\bra{3}\ket{4}}={\bra{4}\ket{1}}(=x_1)$  and  ${\bra{1}\ket{3}}={\bra{2}\ket{4}}(=x_2)$. Hence the expression for the ratio of $D_{total}/S_{total}$ reduces to
\begin{eqnarray}
\frac{D_{total}}{S_{total}}=\frac{1}{4e} \frac {x_1 \sum_{\gamma=1}^{4} \vert V_{\gamma}-V_{\gamma+1}\vert^2+ x_2 \sum_{\gamma=1}^{2} \vert V_{\gamma}-V_{\gamma+2}\vert^2}{x_1 \sum_{\gamma=1}^{4} \vert V_{\gamma}-V_{\gamma+1}\vert+ x_2 \sum_{\gamma=1}^{2} \vert V_{\gamma}-V_{\gamma+2}\vert}.\nonumber\\ 
\end{eqnarray}
In the above expression the sum $ \sum_{\gamma=1}^{4}$ includes a term  $\vert V_{4}-V_{4+1}\vert^2$ where one should identify $V_{4+1}$ with $V_{1}$. 
\\
From the above expression it clear that there are two ways to obtain a universal ratio, i.e. either $(a)$ we have $``x_1\neq0$ and $x_2=0"$ , ``$x_1=0$ and $x_2\neq0$",  ``$x_1=x_2$"  or $(b)$ we tune the voltage differences such that $V_1=V_3$ and $V_2=V_4$. Note that the former condition for obtaining a universal ratio with the already existing constraint of ${\cal Z}_4$ symmetry puts further constraints on the $S$-matrix while the latter  does not put any further constraint on the $S$-matrix though  it does restrict the possible bias voltages that one could apply. The above analysis performed for $n=4$ case with a ${\cal Z}_4$ symmetric $S$-matrix can be straightforwardly extended  to  $n>4$ cases following the geometric way of identifying distinct subsets of overlaps where each element of the subset have the same value owing to the  ${\cal Z}_n$ symmetry. The identification of such overlaps obtained from a geometric approach can be seen in Fig.~\ref{symmetry} c), d) for the case of $n=5$ and $n=6$. It is interesting to note that, for both $n=4$ and $n=5$ we have exactly the same number (in this case, two) of distinct subsets of overlap while for the case of $n=6$ there are three distinct subsets of overlap.
In general for a given case with $n$ number of vertices, the total number of independent subsets of overlaps is $n/2$  for even $n$ and $(n-1)/2$ for odd $n$. One can easily check this by noting the fact that independent subset which are closed under the action of ${\cal Z}_n$ symmetry can be identified as a collection of bonds drawn between vertices on the $n$ sided polygon that are {\cal (a)}  the nearest neighbours  {\cal (b)} the next nearest neighbours  {\cal (c)} the next to next nearest neighbours and so on and so forth respectively. 
Thus, using this observation we arrive at the following expression for the $D_{total}/ S_{total}$ for any given $n$
\begin{eqnarray}
\frac{D_{total}}{S_{total}}= \frac{1}{4e} \frac {\sum_{i=1}^{n_{e}\hspace{0.04cm}/\hspace{0.04cm}n_{o}}   x_{i} \sum_{\langle\gamma,\alpha\rangle}^{i} \vert V_{\gamma}-V_{\alpha}\vert^2}{\sum_{i=1}^{n_{e}\hspace{0.04cm}/\hspace{0.04cm}n_{o}}   x_{i} \sum_{\langle\gamma,\alpha\rangle}^{i} \vert V_{\gamma}-V_{\alpha}\vert},\nonumber\\
\end{eqnarray}
where $x_{i}$ represents the value of the overlap corresponding to $i^{th}$ subset which is a collection of bonds drawn between vertices which are $i^{th}$ nearest neighbour identified in a clockwise sense on the polygon; $n_{e}\hspace{0.04cm}/\hspace{0.04cm}n_{o}$ represent maximum number of possible subsets for an even $n$ given by $n_e=n/2$ or for an odd $n$ given by $n_o=(n-1)/2$. Similarly $\sum_{\langle\gamma,\alpha\rangle}^{i}$ represents sum over all $i^{th}$ nearest neighbour.  It is clear from the above expression that, if we want to have a ratio $D_{total}/S_{total}$ which is independent of the S-matrix then we should either have all the  $x_{i}$'s equal or we can let a few of them to be equal while the rest are set to zero. \\
Note that for the case of an even no. of $n$ edge state we have ($n/2$)  $x_{i}$ variables which needs to be tuned appropriately for obtaining a universal ratio of ${D_{total}}$ to ${S_{total}}$.  This in indeed a considerable reduction in terms of the number of free parameters which needs to be tuned to obtain a constrained space of parameters in which the condition for the universal ratio holds. The $M$-matrix has $n^2-(2n-1)$ free real parameters as the M-matrix itself is formed by replacing each element of the $S$-matrix (which is $(n\cross n)$ unitary matrix) by its respective squared modulus. This can be evaluated by considering the fact that $(n\cross n)$ unitary matrices can be parametrized in terms of $n^2$ real parameters of which $(2n-1)$ parameters\cite{Jarlskog}  are overall phases which go away due to mod operation. Hence in general we will have to tune  $n^2-(2n-1)=(n-1)^2$ real parameters to obtain the subspace of universal ratios. But due to the ${\cal Z}_n$ symmetry of the $M$-matrix this is further reduced to tuning of only  $n/2$ number of $x_{i}$'s, which is a considerable reduction from the original parameter space of $(n-1)^2$ parameters. 

\noindent{{\underline{Discussion and Conclusions}:-}} In this article we have discussed power dissipation in a quantum Hall circuitry and presented an analysis of dissipation due to a quantum scatterer like a QPC. We have pointed out that the total dissipation in the circuit can be understood as a sum of dissipation at the ``contact between the edge state and the electron reservoir" and ``the dissipation at the QPC". The main focus of the article was to analyse the dissipation at the QPC alone which we called $D_{total}$. We also discussed a subdivision of $D_{total}$ into its components $D_{i}$ which can be understood as power dissipation happening in the individual outgoing edges emanating from the QPC region. Then we go on to  study the ratio of the total dissipation induced by the QPC which is scattering electrons between $n$ chiral edge channels of $\nu=1$ QH state to the corresponding total auto-correlated noise generated in the outgoing edge channels given by $D_{total}/S_{total}$. This can be thought of as an empirical out of equilibrium fluctuation-dissipation type relation for free chiral fermions. It is shown that $D_{total}/S_{total}{\small{(}}\omega=0{\small{)}} = V/{\small{(}}4  e{\small{)}}$, where $e$ is the electronic charge and $V$ is the voltage imposed on one of the ``$n$'' incoming edge channels while all other ``$n-1$'' edge channels are kept grounded,  which is universal except for the linear voltage dependence.\\
We further show that this ratio can also be universal except for a function of bias voltages provided the corresponding $S$-matrix respects ${\cal Z}_n$ symmetry (symmetric under cyclic permutation of edge channels) when distinct bias is imposed on each of the edge state. \\
For  the case of $n=3$, we show that ${\cal Z}_3$  symmetry is enough to ensure a universal ratio but for $n>3$ further constraints need to be imposed. We provide a purely geometric visualization of the consequence of imposition of the ${\cal Z}_n$ symmetry and additional constraints that need to be imposed on the $M$-matrix which leads to a universal $D_{total}/S_{total}$ ratio. We have also obtained a closed form expression for} $D_{total}/S_{total}$ for a general $n$ which is one of the central results of this article. \\
Also, note that $S_{total}$ is a readily measurable quantity in QH experiments and  we have shown that $D_{total}$ is also an independently measurable quantity via a voltage probe measurement. Hence the  universal ratio of $D_{total}$ to $S_{total}$ can be thought of as a measurable quantity. This fact opens up an interesting aspect for experimental exploration of possibilities for observation of deviation from this empirical relation which could be considered as an indicator of non Fermi-liquid (Luttinger liquid) owing to the fact that our results are derived assuming a Fermi-distribution function for electrons on the edge.  Also it should be noted that most of theoretically proposed  and experimentally implemented probes for search of non-Fermi liquid behaviour for QH edge states involves tunneling current measurement\cite{Roddaro} or shot noise measurement\cite{Hashisaka}. Measurement of our empirical relation  could provide yet another independent route for quantification of a deviation from Fermi liquid behaviour which is different from both of these approaches but fuses elements of both (a combination of both noise and conductance measurement).\\
Lastly it is important to note that our results crucially  depends on the assumption that the $S$-matrix elements are independent of energy of the incident electron. Such an assumption is always valid in bias window which is small enough not to explore the nonlinear energy dependence of the density of the states of either the scattering region and the leads. This automatically implies that must not have transport resonances in the bias window. \\

\noindent{{ \underline{Acknowledgements}:-}} We acknowledge discussions with Eytan Grosfeld during early stages of the project. SD thanks Anindya Das and Biswajit Karmakar for numerous discussions on quantum Hall edge states and possible measurement related to our proposal. DW thanks Ora Entin Wohlman for her remarks on the work.

\bibliographystyle{apsrev}

\bibliography{references}

\end{document}